# Pricing Global Macroeconomic Risk in Equity Markets: Evidence from Selected G20 Economies

Vivek Mishra
Department of Pure and Applied Mathematics
Alliance University
Bengaluru, India
vivek.mishra@alliance.edu.in

*Abstract*— This study investigates whether international equity markets systematically price global macroeconomic risks. The empirical analysis is conducted using monthly excess returns for ten G20 countries over the period 2000–2024. A Dynamic Factor Model (DFM) is employed to extract latent global factors from a set of macroeconomic variables capturing global inflation, real activity, monetary policy, term structure, exchange rates, volatility, and oil prices. The model selection criteria of the dynamic factor framework, which support a 3-factor specification that is parsimonious. The Fama–MacBeth regressions demonstrate the low explanatory power of the 3-factor model. In contrast, a 4-factor specification results in economically large and statistically significant factor loadings, an obvious rise in explanatory power, and a significant improvement in model performance. The results indicate that a four-factor specification provides the best balance between explanatory power and model stability, significantly improving the ability to explain cross-sectional variation in excess returns ($R^2 \approx 0.22$), with all factors statistically significant. The Capital Asset Pricing Model, while offering a parsimonious and stable benchmark with consistently significant market betas, exhibits limited explanatory power due to its single-factor structure. Overall, the findings suggest that macro driven latent factors extracted through the DFM provide a more comprehensive and empirically robust framework for international asset pricing than the CAPM, highlighting the importance of incorporating multiple sources of systematic risk in explaining cross-country equity returns.

## INTRODUCTION

The models used in classical factor analysis divide the observed vector into two components: the unobserved systematic part and the unobserved errors. While the systematic portion is regarded as a linear combination of a comparatively limited number of unobserved factors, the components of the error vector are regarded as uncorrelated or independent. The factor analysis separates the effect of factors to the effect of errors. When the objective is to summarize high-dimensional data or find hidden common shocks, models that consider factors as latent random variables retrieved from the covariance structure like Principal Component Analysis (PCA) or dynamic factor models (DFM) are frequently more suitable in statistical finance and classical factor analysis.

The approaches in the factor model are the three-factor model of Eugene F. Fama, Kenneth R. French [1]) where they have considered the three factors as an overall market factor, factors related to firm size and book-to-market equity. The five-factor model proposed by Eugene F. Fama, Kenneth R. French [2] where they have added two more factors of profitability and the Investment factor. Rahul Roy and Santhakumar Shijin [3] introduce a six-factor model by introducing the sixth factor as human capital to Fama and French five factor model.

In the Capital assets pricing model (CAPM), it is expected that excess returns are proportional to the covariance of assets with its market portfolio, but CAPM fails to capture cross sectional patterns as shown by Basu, S. [4]. Fama and French model has gained a wide support empirically, but it also could not explain the market anomalies like Momentum pointed out by Jegadeesh, N., & Titman, S. [5]. Merton, R. C. [6]. has proposed a new way to think about the risk and return. As for the Merton R.C., the factor loading and the factors should vary over time depending on the economic conditions. Harvey, C. R. [7] proposed the time varying the conditional covariance in the tests of the asset pricing models. Shanken[8]investigates whether the Intertemporal Capital Asset Pricing Model (ICAPM), initially proposed by Merton [6], is supported by real financial market data where Merton has considered that the asset return does not only depend on the market risk but also on the Hedging risks related to future changes in the investment opportunity set.

The Fama-French method is developed with the thought that asset return depends on the ad-hoc, predetermined microlevel variables rather than systemic macroeconomic determinants of asset returns. This becomes a major point of criticism for Fama French model. Parallelly a body of scholars concentrates on identifying latent macroeconomic drivers within a broad panel of macro-financial indicators. Stock, J. H., & Watson, M. W. [9] used PCA based factor extraction for forecasting in large data sets. Stock, J. H., & Watson, M. W. [10] also proposed the diffusion index methodology for macroeconomic forecasting and evaluating its performance. Ludvigson and Ng [11], [12] created a macro-finance framework that uses Principal Component based Dynamic Factor Models (DFM) to extract a few latent common factors from a wide range of macroeconomic variables, including industrial production, labour market indicators, credit conditions, inflation, interest rates, and financial-market variables.

It is demonstrated that these estimated macro parameters have significant predictive potential for excess stock returns and bond risk premia. It outperforms more conventional forecasting variables like yield spreads or dividend yields. The Generalized Dynamic Factor Model (GDFM), created by Forni, Hallin, Lippi, and Reichlin [13], [14], provides an effective framework for examining huge macroeconomic datasets. According to Adrian, Etula, and Muir's [15] macro financial asset pricing model, the ability of financial intermediaries, especially broker-dealers, to bear risk is crucial to understanding and predicting asset returns. Doz, C., Giannone, D., & Reichlin, L. [16] proposes a practical two-step estimator of PCA initialization and Kalman smoother for

DFMs that allows for factor dynamics and heteroskedastic idiosyncratic errors widely adopted in applied work.

Doz, C., Giannone, D., & Reichlin, L. [17] develops QML estimation methods and theoretical properties for the two-step Kalman-filter approach; important for reliable inference with DFM. Jungbacker, B., & Koopman, S. J. [18] developed likelihood-based estimation, inference and computational algorithms for DFM. According to Barigozzi M. & Luciani M. [19], contemporary Dynamic Factor Models (DFMs) can permit several layers of factors rather than just one common factor. They also talk about regularized or sparse DFMs, which manage big macro financial datasets by retaining only the most significant factor loadings. They also compared time-domain methods with frequency-domain methods that differentiate long-term trends from short-term cycles.

Alessandro Giovannelli, Daniele Massacci, Stefano Soccorsi [20] compared the static and generalized dynamic factor model and shown that dynamic factor model is more accurate. Karen Miranda, Pilar Poncela & Esther Ruiz [21] concludes that researchers must be careful in choosing the number of factors, estimation method, and dynamic structure, as the wrong specification can lead to misleading economic conclusions and poor predictive performance. Matteo Barigozzia, Filippo Pellegrino [22] expands the classical. DFM into a multidimensional framework that can handle several groups of factors and large panels of time series, capturing complex dependence structures in macro–financial data. Omer Faruk Akbal [23] finds that a regime-switching dynamic factor model estimated via an Expectation Maximization (EM) algorithm delivers a strong trade-off between accuracy and computational speed offering robust nowcasting of macroeconomic states and successfully matching historical recession dates even with large panels of data.

The main objective of this study is to find out if international equity markets systematically price global macroeconomic risks. The analysis compares an observable, economically based macro-factor framework with a latent Dynamic Factor Model (DFM) as alternate methods for capturing global macroeconomic dynamics to answer this question. The study assesses the pricing relevance, stability, and economic interpretability of these factor structures within a cross-sectional asset pricing framework using a panel of country-level equity returns and macroeconomic indicators from 10 economies of G20.

## I. DATA DESCRIPTION AND SAMPLE SELECTION

This study uses panel data on equity market returns for ten of the G-20 economies from January 1, 2000, to December 31, 2024, United States, United Kingdom, Japan, Italy, India, Germany, France, China, Canada and Brazil. The purpose of considering these countries as these countries are the major economies of G-20. These countries represent the developed economies and the developing economies a balanced mixture of these economies is considered, which approximately represents 60-65% of the world's economies. To represent national equity markets, we use the primary benchmark index for each country: Brazil's Bovespa, Canada's S&P/TSX Composite, China's SSE Composite, France's CAC 40, Germany's DAX, India's BSE Sensex, Italy's FTSE MIB, Japan's Nikkei 225, the United Kingdom's FTSE 100, and the United States' S&P 500.

Broad market indexes are used to represent each nation's equity market performance, guaranteeing uniformity and comparability across global markets. Yahoo Finance[24] is the source of all market indices. The sample period provides an appropriate framework for evaluating frequent global shocks affecting equity returns since it encompasses several economic regimes, including the dotcom bubble, the global financial crisis, the COVID-19 pandemic, and the post-pandemic recovery. All index price series are converted to monthly frequency and transformed into logarithmic returns to enhance stationarity and comparability across countries. To account for the time value of money and to adopt a global investment perspective, returns are expressed in excess form by subtracting a global risk-free rate, proxied by the U.S. three-month Treasury bill rate. If $R_{f,t}$ is monthly simple risk-free rate in decimal form and $y_t$ is the annualized 3-month Treasury bill yield in percent then $R_{f,t} = \frac{y_t}{100 \times 12}$ and Log Monthly risk-free rate= $\ln(1 + R_{f,t})$ and $R_{i,t}$ is the log of the return where $R_{i,t} = \ln\left(\frac{Index\ level\ this\ mon}{Index\ level\ last\ month}\right)$ hence the excess return is found as

$$excess\ return = \ln(1 + R_{i,t}) - \ln(1 + R_{f,t}) \quad \ldots (1)$$

the logarithmic monthly risk-free rate, which is derived from the yield on three-month US Treasury bills obtained from the Federal Reserve Economic Data (FRED).

Inflation, Industrial Production, Monetary Policy rates, Term Spreads, and Exchange rate returns are among the macroeconomic and financial variables employed in in this study. The log difference of the Consumer Price Index is used to calculate inflation. The Industrial Production Index growth rate used to represent economic activity. Short-term policy interest rates are used to measure monetary policy, and they are converted into first differences to guarantee stationarity. The difference between the yields on long-term government bonds and the short-term policy rates is a representation of the term structure of interest rates. Log returns of nominal exchange rates are used to quantify changes in exchange rates.

To guarantee consistency and comparability, the macroeconomic dataset is compiled from globally reputable sources. The World Bank's World Development Indicators database is used to retrieve consumer price index (CPI)[25]. The International Monetary Fund's International Financial Statistics database provides information on bilateral exchange rates, long-term government bond yields, industrial production indices, and monetary policy rates[26]. The Volatility Index (VIX)[24], which measures overall market volatility, serves as a proxy for the state of the world economy is accessed from yahoo finance. To account for global commodity price shocks, crude oil prices are included as the sole global macroeconomic factor, measured as log returns[28].To prepare the macroeconomic and financial series for analysis all the series are transformed in stationary series either by taking the log difference( for growth rate) or by first difference(for level change) as shown in the table no.1

*Table 1:Summary of Variables and Transformations*

| Variable | Transformation | Units/ Description |
|---|---|---|
| Inflation (CPI) | $100\Delta log(CPI_t)$ | Monthly % change in prices |
| Industrial Production (IP) | $100\Delta log(IP_t)$ | Monthly % growth |
| Monetary Policy | $\Delta r_t = r_t - r_{t-1}$ | Monthly change in policy rate (percentage points) |
| Government Bond Yield | $\Delta y_t = y_t - y_{t-1}$ | Monthly change in bond yields (percentage points) |
| Exchange Rate (FX) | $100\Delta log(FX_t)$ | Monthly % change in FX rate |
| Volatility | $100\Delta log(V_t)$ | Monthly % change in realized volatility |
| WTI Crude Oil Price | $100\Delta log(WTI_t)$ | Monthly % growth in oil price |

## II. Methodology

The majority of macroeconomic datasets are made up of numerous financial and economic series that show significant co-movement across time. By utilizing a limited number of unobserved latent factors to capture the joint dynamics of numerous observed variables, Dynamic Factor Models (DFMs) offer a framework for summarizing this data. While idiosyncratic components record variable-specific movements, these elements are understood as common economic causes, causing oscillations throughout the economy. To estimate the latent factors the factor analytical model is used.

$$X_t = ZF_t + e_t \quad …(2)$$

Where $X_t$ is the vector of M-dimensional vector of variable which is observed and $F_t$ is the Nx1 variables of the unobserved factors. $Z$ is called the factor loading which will be the matrix having order M x N. $e_t$ is the error term where $e_t \sim \mathcal{N}(0, R)$. This model is static in the sense that it does not consider the autocorrelation of the factors into consideration so in DFM has the second stage also which is called its transition state.

$$F_t = TF_{t-1} + \eta_t \quad …(3)$$

Where $T$ is the transition matrix which governs the persistence of the common factor and $\eta_t \sim \mathcal{N}(0, Q)$ is a factor innovation. Equation 2,3 is a state-space representation of the DFM. The panel data has been made and by using Kalman filter, the latent factor, factor loadings, and dynamic parameters via maximum likelihood have been estimated.

### A. Kalman Filter

The prediction in the Kalman filter about the factor at the instant 't' is represented as $a_t = E[F_t/X_{1:t}]$. The predicted value of the factor one step ahead using the data till t-1 is $a_{t/t-1} = E[F_t/X_{1:t-1}]$. The filtered covariance matrix is represented as $P_t = Var[F_t/X_{1:t}]$ and the predicted covariance matrix is $P_{t/t-1} = Var[F_t/X_{1:t-1}]$. Kalman filter works in two steps with recursion. The first step is prediction step as in equation 4 and 5

$$a_{t|t-1} = T\, a_{t-1} \quad …(4)$$
$$P_{t|t-1} = TP_{t-1}T' + Q \quad …(5)$$

the predicted covariance is calculated in equation 5 with the noise Q. The second step in the Kalman filter is the update step where the prediction error $v_t$ is calculated as $v_t = X_t - Za_{t/t-1}$ with the variance of the prediction error as $S_t$ where

$$S_t = Z\, P_{t/t-1}Z' + R \quad …(6)$$

and R is the noise. Kalman gain is the weightage assigned to the new information relative to prior prediction of the factors value is calculated as $K_t = P_{t/t-1}Z'S_t^{-1}$. The factor estimate can be updated as $a_t = a_{t|t-1} + K_t\, v_t$ and the updated covariance matrix is $P_t = P_{t|t-1} - K_tS_tK_t'$. Under the assumption of Gaussian errors of prediction, the sequence of prediction errors $v_t$ is normally distributed with mean zero and covariance matrix $S_t$. The log-density of the prediction error is added over time to create the log-likelihood function of the model parameters, which is provided by:

$$\mathcal{L}(\theta) = -\frac{1}{2}\sum_{t=1}^{T}\left[log|S_t| + v_t'S_t^{-1}v_t + Mlog(2\pi)\right] \quad …(7)$$

The maximum likelihood estimates of the model parameters are obtained by maximising this likelihood function. In this case, M represents the dimension of the vector, $X_t$, or the number of observed variables in the measurement equation. This approach offers advantages like allowing time varying latent factors, handling missing observations and estimating large macroeconomic panels. Negative realisations of the factor indicate worldwide downturns or accommodating circumstances, while positive realizations correspond to times of global economic expansion or tightness. The degree to which each macroeconomic variable reacts to shocks from around the world is revealed by factor loadings.

### B. Fama–MacBeth Two-Pass Regression Methodology

The Fama MacBeth [29] two-pass regression methodology is to examine whether systematic risk factors are priced in the cross-section of asset returns and to estimate the corresponding risk premia. The approach is particularly suitable for asset pricing, as it separates the estimation of factor exposures from the estimation of factor prices while allowing for time-varying cross-sectional relationships. In the first pass of the Fama–MacBeth procedure, excess returns for each asset $i$ are regressed on a set of $K$ risk factors over time to estimate factor loadings (betas). Formally, for asset $i = 1…N$, and time period $t = 1…T$, the model is given by

$$R_{i,t}^e = \alpha_i + \beta_i^T F_t + \epsilon_{i,t} \quad …(8)$$

Where $R_{i,t}^e$ denotes the excess return of asset $i$ at time $t$, $\alpha_i$ is the asset-specific intercept, $F_t = (F_{1,t}, F_{2,t}……F_{K,t})^T$ is a $K \times 1$ vector of estimated latent macroeconomic factors obtained from the DFM, $\beta_i = (\beta_{i,1}, \beta_{i,2}……\beta_{i,K})^T$ represents the factor loadings (betas) for asset $i$, and $\epsilon_{i,t}$ is the idiosyncratic error term. The factor loadings $\hat{\beta}_i$ are estimated from time-series regressions over the full sample. These estimated betas are then used in a cross-sectional regression of average excess returns to estimate the corresponding factor risk premia.

In the second pass of the Fama MacBeth methodology, the estimated factor loadings $\hat{\beta}_i$ from the first-pass time-series regressions are used to explain the cross-section of asset returns at each point in time. For each time $t$, the following cross-sectional regression is estimated

$$R_{i,t}^e = \lambda_{0,t} + \hat{\beta_i^T}\lambda_t + u_{i,t}\ , i = 1,2,3……N \quad …(9)$$

Where $\hat{\beta_i^T}$ is the estimated Beta from the first pass, $\lambda_t = (\lambda_{1,t}, \lambda_{2,t}……\lambda_{K,t})^T$ time t factor risk premia, $\lambda_{0,t}$ cross sectional intercept, $u_{i,t}$ pricing error. This regression is estimated separately for each time period $t$, producing a time series of factor risk premia $\{\lambda_t\}_{t=1}^T$. The average factor risk

premia are then computed as the time-series averages of the period-specific estimates

$$\bar{\lambda} = \frac{1}{T}\sum_{t=1}^{T}\lambda_t \quad \ldots (10)$$

Statistical inference on $\bar{\lambda}$ is conducted using heteroskedasticity and autocorrelation consistent (HAC) standard errors to account for time series dependence in $\lambda_t$.

## III. RESULT AND DISCUSION

To create the panel of data the series of the consumer price index is considered (CPI). The stationarity of the monthly inflation was initially examined using the Augmented Dickey–Fuller (ADF) test and the Kwiatkowski-Phillips-Schmidt-Shin (KPSS) Test. While both tests indicate stationarity for most countries, conflicting results were observed for France, India, and Japan. Specifically, the ADF test failed to reject the null hypothesis of a unit root, whereas the KPSS test suggested stationarity. To address this inconsistency and account for potential structural breaks in the series, the Zivot–Andrews Test was applied for these countries. The results confirm stationarity with one structural break in table 2.

*Table 2: Unit Root Test Results for Monthly Inflation*

| Country | ADF p-value | KPSS p-value | Zivot–Andrews Statistic | Zivot–Andrews p-value | Conclusion |
|---|---|---|---|---|---|
| Brazil | 0.0000 | 0.1000 | – | – | Stationary |
| Canada | 0.0069 | 0.1000 | – | – | Stationary |
| China | 0.0030 | 0.1000 | – | – | Stationary |
| France | 0.0999 | 0.1000 | -14.047 | 0.000 | Stationary with structural break |
| Germany | 0.0381 | 0.0430 | – | – | Stationary |
| India | 0.2343 | 0.1000 | -18.734 | 0.000 | Stationary with structural break |
| Italy | 0.0038 | 0.1000 | – | – | Stationary |
| Japan | 0.2324 | 0.0100 | -15.357 | 0.000 | Stationary with structural break |
| United Kingdom | 0.0224 | 0.0100 | – | – | Stationary |
| United States | 0.0224 | 0.0100 | – | – | Stationary |

The second macroeconomic data considered in the panel is Industrial production (IP). The monthly (IP) data is considered from 2000-01-31 to 2024-12-31. The I P data for China is not consistent after the covid period so the series of China have been dropped .The series have been stationaries with the fist difference and due to large number of missing values in the series of France, Germany, Italy these series have been dropped and remaining series of Japan, India, Canada, Brazil, United Kingdom and United states are considered. The third data is of the series Monetary Interest rates where the series is stationary after the first difference the data of the countries Brazil, Canada, France, Germany, Italy, Japan and United States are considered as per the availability. The government bond yield is the next considered data series for the countries United States, Japan, Germany, United Kingdom, Canada, France and Italy. Another series is the monthly money exchange rate for the countries like Brazil, Canada, India, Japan and United Kingdom. The series of France, Germany and Italy have been dropped as the series have majorly missing values after stationarities. Market volatility is proxied by the Chicago Board Options Exchange (CBOE) Volatility Index (VIX). The index reflects market expectations of volatility derived from options on the S&P 500 Index. Commodity price dynamics are captured using the monthly log growth of West Texas Intermediate (WTI) crude oil prices for the period 2000–2024, which serves as a proxy for global commodity price fluctuation.

On concatenating all these data, a panel of data has been created with the shape is (300,7) the descriptive statistics of this panel are represented in table 3.

*Table 3: Descriptive statistics of the panel data*

| Variable | Mean | Std Dev | Min | Max |
|---|---|---|---|---|
| Brazil Inflation | 0.497 | 0.392 | -0.682 | 2.976 |
| Canada Inflation | 0.184 | 0.388 | -1.043 | 1.420 |
| China Inflation | 0.165 | 0.602 | -1.399 | 2.567 |
| France Inflation | 0.139 | 0.344 | -1.247 | 1.413 |
| Germany Inflation | 0.157 | 0.385 | -1.036 | 1.962 |
| India Inflation | 0.495 | 0.729 | -1.660 | 4.474 |
| Industrial Production (India) | 1.871 | 2.827 | -6.102 | 9.183 |
| Industrial Production (Brazil) | 0.079 | 2.373 | -21.946 | 11.659 |
| Interest Rate (US) | -0.002 | 0.163 | -0.973 | 0.450 |
| Interest Rate (Japan) | -0.001 | 0.035 | -0.250 | 0.350 |
| US Bond Yield Diff | -0.008 | 0.218 | -1.110 | 0.650 |
| Germany Bond Yield Diff | -0.011 | 0.162 | -0.530 | 0.761 |
| UK Bond Yield Diff | -0.005 | 0.187 | -0.693 | 1.173 |
| Brazil FX | -0.405 | 4.834 | -25.371 | 14.874 |
| India FX | -0.223 | 1.573 | -6.566 | 4.359 |
| Japan FX | -0.138 | 2.687 | -8.501 | 7.658 |
| UK FX | -0.092 | 2.478 | -10.686 | 8.586 |
| VIX Volatility | -0.142 | 21.385 | -61.428 | 85.259 |
| WTI Oil Growth | 0.355 | 10.918 | -78.196 | 61.503 |

This data is further standardized and then the dynamic factor modelling (DFM) have been implemented on it with the factors K=1,2,3,4,5,6,7 and the factor order 1.The Akaike Information Criterion (AIC) and the Bayesian Information Criterion (BIC) criteria are calculated with respect to every factor number as represented in table 4.

*Table 4: Comparison of AIC and BIC across different factor specifications*

| No of factor(K) | AIC | BIC | Condition Number |
|---|---|---|---|
| 1 | 29604 | 29882 | 1.0e+20 |
| 2 | 29066 | 29492 | 1.0e+19 |
| 3 | 20882 | 21463 | 1.0e+28 |
| 4 | 21208 | 21953 | 1.0e+27 |
| 5 | 20494 | 21409 | 1.0e+27 |

| | 6 | 20856 | 21950 | 1.0e+28 |
|---|---|---|---|---|
| | 7 | 21106 | 22384 | 5.0e+28 |

Note: (*The condition number is to assess the numerical stability of estimated covariance matrix. Higher values indicate the potential multicollinearity and instability in parameter estimates*)

From table 4, both the AIC and the BIC attain their minimum values at K = 5, suggesting that a five-factor model provides the best fit among the specifications considered. AIC values reveal a substantial decline when moving from K = 1 to K = 2, followed by an even sharper reduction between K = 2 and K = 3, indicating significant improvement in model fit with the inclusion of additional factors. However, beyond K = 3, the rate of decline in AIC diminishes considerably, suggesting diminishing marginal gains from adding more factors. Furthermore, it is observed that the condition number of the estimated covariance matrix increases rapidly as the number of factors rises, indicating potential numerical instability in higher-dimensional models.

While model selection criteria in the Dynamic factor modelling framework favors a parsimonious 5 factor model but the asset pricing results tell a different story. The Fama-Macbeth regressions which are done with K=3,4,5 show that K=3 model has limited explanatory power with the statistical insignificant risk premia and negligible $R^2$. In contrast K=4 model substantially improves the model performance yielding the economically large and statistically significant factor loading along with a marked increase in the explanatory power. This divergence suggests that factors relevant for capturing the macroeconomic co-movement may not be sufficient for explaining cross sectional variation in asset returns. This distinction highlights the tradeoff between Statistical parsimony and asset pricing performance. The Fama-Macbeth regressions result for K=3,4 and 5 are presented in table 5

*Table 5: Fama–MacBeth Regression Results Highlighting the Preferred Four-Factor Specification*

| Variables | K=3 | K=4 | K=5 |
|---|---|---|---|
| Observations | 2990 | 2990 | 2990 |
| Intercept | 0.0019(0.0026) | 0.0020(0.0020) | 0.0020(0.0021) |
| Factor 1 | 0.0225(0.0425) | -0.3014(0.0397) | -0.1060(0.0351) |
| Factor 2 | -0.0033(0.0155) | 0.1693(0.0175) | 0.0188(0.0105) |
| Factor 3 | -0.0180(0.0160) | -0.1031(0.0123) | 0.0083(0.0123) |
| Factor 4 | - | -0.0676(0.0065) | -0.0158(0.0031) |
| Factor 5 | - | - | -0.0158(0.0031) |
| $R^2$ | 0.016 | 0.220 | 0.216 |

Note: *The standard error (in parentheses)*

This result shows that the three-factor model provide the explanatory power with $R^2$ of 1.65 in contrast the four-factor model exhibits a substantial improvement in the explanatory power with $R^2$ value as 22% and all factors risk premia statistically significant at conventional level. Expanding the model to five factors does not lead to further improvement while the overall explanatory power remains similar some factor loading become statistically insignificant suggesting the inclusion of redundant factors and potential overfitting.

To evaluate the performance of traditional asset pricing models against macro-driven latent factor models, the Capital Asset Pricing Model (CAPM) is compared with the Dynamic Factor Model (DFM). The CAPM results as shown in table 6 indicate that the market beta is statistically significant across all countries, confirming the presence of systematic risk. However, the explanatory power varies across countries, with $R^2$ values ranging from approximately 0.09 to 0.65. This suggests that while market risk is important, it does not fully explain cross-sectional variations in excess returns.

*Table 6: CAPM Regression Results for Each Country*

| Country | Alpha | Alpha P Value | Beta | Beta P Value | $R^2$ |
|---|---|---|---|---|---|
| BRAZIL | 0.001938 | 0.548451 | 0.970369 | 2.070821e-32 | 0.377384 |
| CANADA | -0.000389 | 0.781964 | 0.740115 | 1.064020e-69 | 0.650136 |
| CHINA | -0.000434 | 0.911694 | 0.486563 | 6.114860e-08 | 0.094174 |
| FRANCE | -0.003724 | 0.032613 | 0.926618 | 6.727920e-71 | 0.656569 |
| GERMANY | -0.001459 | 0.473921 | 1.061175 | 6.778485e-69 | 0.645754 |
| INDIA | 0.004933 | 0.100919 | 0.792812 | 1.348556e-26 | 0.319029 |
| ITALY | -0.005351 | 0.030157 | 0.963611 | 1.311160e-47 | 0.507620 |
| JAPAN | -0.001764 | 0.462342 | 0.798898 | 9.460042e-38 | 0.426629 |
| UK | -0.002894 | 0.039135 | 0.682445 | 1.429137e-63 | 0.615332 |

The Dynamic Factor Model incorporates latent factors extracted from macroeconomic variables. The Fama–MacBeth regression results demonstrate that the four-factor specification delivers consistent and economically meaningful explanatory power, with an $R^2$ of approximately 0.22 and all factors statistically significant. This indicates that macroeconomic dynamics play a crucial role in explaining asset returns beyond the market factor.

The results as in table 7 demonstrate that the DFM with four factors, outperforms the CAPM by capturing additional sources of systematic variation in returns. However, this improvement comes at the cost of reduced numerical stability, highlighting a trade-off between model complexity and robustness. Although the factors in the Dynamic Factor Model are latent, their economic meaning can be inferred from the underlying macroeconomic variables used in the analysis. These interpretations suggest that asset returns are influenced by multiple macroeconomic channels, supporting the use of a multi-factor framework over the single-factor CAPM.

*Table 7: Evidence on the Relative Performance of CAPM and Dynamic Factor Models*

| Model | Factors | Avg. $R^2$ | Significant Factors | Stability | Interpretation |
|---|---|---|---|---|---|
| CAPM | 1 (Market) | ~0.45 (varies by country) | Beta significant in all cases | Stable | Captures market risk only |
| DFM (k=3) | 3 latent factors | 0.016 | None significant | Unstable | Underfitting |
| DFM (k=4) | 4 latent factors | **0.220** | All significant | Moderate instability | Best balance |
| DFM (k=5) | 5 latent factors | 0.216 | Most significant | Highly unstable | Overfitting risk |

## CONCLUSION

The current research shows that when it comes to explaining excess returns across G20 countries, the Dynamic Factor Model (DFM) offers a more reliable framework than the Capital Asset Pricing Model. In addition to the AIC and BIC criteria, Fama-MacBeth regression results are used to validate the optimal number of factors. The findings suggest that while a parsimonious 3-factor structure may capture macroeconomic dynamics, an additional factor is necessary to adequately explain cross-sectional variation in asset returns so a 4-factor model is used.The findings demonstrate that the CAPM's single-factor structure restricts its capacity to capture more extensive sources of systematic risk, even though it offers a reliable and frugal benchmark with consistently significant market betas. The DFM, on the other hand, shows better explanatory power especially the four-factor specification. Although higher-dimensional specifications introduce numerical instability, the evidence suggests that incorporating macro-driven latent factors offers a more comprehensive understanding of asset return dynamics. Overall, the results acknowledge the simplicity and robustness of the CAPM as a baseline and support the use of multi-factor frameworks over it.